\documentclass[12pt]{spieman}  
\usepackage{amsmath,amsfonts,amssymb}
\usepackage{graphicx}
\usepackage{setspace}
\usepackage{tocloft}

\usepackage{soul}
\usepackage{algorithm}
\usepackage{algorithmic}

\title{An Exploration of Blockchain Enabled Decentralized Capability based Access Control Strategy for Space Situation Awareness}

\author[a]{Ronghua Xu}
\author[a*]{Yu Chen}
\author[b]{Erik Blasch}
\author[c]{Genshe Chen}
\affil[a]{Department of Electrical and Computer Engineering, Binghamton University, SUNY, Binghamton, NY, USA, 13902}
\affil[b]{The U.S. Air Force Research Lab, Rome, NY, USA, 13441}
\affil[c]{Intelligent Fusion Technology, Inc., Germantown, MD, USA, 20876}

\cftpagenumbersoff{figure}
\cftpagenumbersoff{table} 
\begin{document} 
\maketitle

\begin{abstract}
Space situation awareness (SSA) includes tracking of active and inactive resident space objects (RSOs) and assessing the space environment through sensor data collection and processing. To enhance SSA, the dynamic data-driven applications systems (DDDAS) framework couples on-line data with off-line models to enhance system performance. Using feedback control, sensor management, and communications reliability. For information management, there is a need for identity authentication and access control to ensure the integrity of exchanged data as well as to grant authorized entities access right to data and services. Due to decentralization and heterogeneity of SSA systems, it is challenging to build an efficient centralized access control system, which could either be a performance bottleneck or the single point of failure. Inspired by the blockchain and smart contract technology, this paper introduces BlendCAC, a decentralized authentication and capability-based access control mechanism to enable effective protection for devices, services and information in SSA networks. To achieve secure identity authentication, the BlendCAC leverages the blockchain to create virtual trust zones, in which distributed components could identify and update each other in a trustless network environment. A robust identity-based capability token management strategy is proposed, which takes advantage of the smart contract for registration, propagation, and revocation of the access authorization. A proof-of-concept prototype has been implemented on both resources-constrained devices (i.e., Raspberry PI nodes emulating satellites with sensor observations) and more powerful computing devices (i.e., laptops emulating a ground network), and is tested on a private Ethereum blockchain network. The experimental results demonstrate the feasibility of the BlendCAC scheme to offer a decentralized, scalable, lightweight and fine-grained access control solution for space system towards SSA.
\end{abstract}


{\noindent \footnotesize\textbf{*}Correspondence:  \linkable{ychen@binghamton.edu} }

\begin{spacing}{2}   

\section{Introduction}
\label{sect:intro}  
Recent advances in Big Data (BD) have focused research on the volume, velocity, veracity, variety, and value of dynamic data. These developments enable new opportunities in information management, visualization, machine learning, and information fusion that have potential implications for space situational awareness (SSA) \cite{blasch2017big}. In SSA systems, the space environmental data can be collected and processed to determine object motions and models updates \cite{wheeler2016satellite,teehan2007responsive}. A common example is space tracking using electro-optical sensors \cite{jia2016cooperative}. The key aspect for SSA is to track the many resident space objects (RSO), either satellites, debris, or space transportation support \cite{oliva2013applying}. To enable the space environment assessment, many attributes are considered, including communications, space weather, and conjunction analysis as well as using non-tradition data \cite{blasch2018dddas}. As one of the most critical research areas in SSA, there is a need for network management of the space surveillance network \cite{chin2009game, kennewell2013overview}. One development for resource management includes the dynamic data-driven applications system (DDDAS) framework whereby measurements are injected into the execution model to enhance system performance. A DDDAS-based system integrates on-line data with the off-line models creating a positive feedback loop, where the model judiciously guides the sensor selection and data collection, from which the sensor measurements improve the accuracy of the model \cite{blasch2018dddas}.

DDDAS technology can enhance the space network, where components interact and cooperate with each other to build a big data platform to provide a wide range of services \cite{dddas18}. Thus, a huge number of entities, e.g., physical RSOs and virtual services, connect and produce space environment data that can be retrieved by users regardless of their location. To reduce data security risks such as information theft and data alteration, systems dictate that only authenticated and authorized entities are allowed to access the data and use the services provided by the system. The conventional access control approaches have been widely used in the Internet Technology (IT) ecosystem. However, the existing security solutions are not fully adapted to a space network ecosystem due to the constrained resources of space objects and heterogeneity of the platforms. The combination of multiple security technologies and solutions leads to an extraordinary high overload on the system. Furthermore, today's access control solutions often rely on a centralized architecture, which not only demonstrates enormous scalability issues in an distributed environment composed of large number of nodes, but also can be a performance bottleneck or the single point of failure. Consequently, it is necessary to propose new access control solutions for SSA systems.

The blockchain protocol has been recognized as the potential candidate to revolutionize the fundamentals of IT technology because of its many attractive features and characteristics, such as supporting decentralization and anonymity maintenance \cite{crosby2016blockchain}, as well as a fundamental protocol of Bitcoin \cite{nakamoto2008bitcoin}, the first digital currency. In this paper, a BLockchain-ENabled, Decentralized, Capability-based Access Control (BlendCAC) scheme is proposed to enhance the security of space applications. BlendCAC provides a decentralized, scalable, fine-grained and lightweight authentication and access control mechanism to protect devices, services and information in space networks. To achieve secure identity authentication, a decentralized authentication mechanism is implemented on the blockchain, and aims at creating virtual trust zones to allow all distributed entities to identify each other and communicate securely in the trustless network environment. An identity-based capability token management strategy is presented and the federated authorization delegation mechanism is illustrated. A proof-of-concept prototype has been developed and evaluated on a private Ethereum blockchain network, and the experimental results demonstrate the feasibility and effectiveness of the proposed BlendCAC scheme.

The major contributions of this work are as follows:
\begin{enumerate}
\item Leveraging the blockchain and smart contract technologies, a decentralized access control solution is proposed to address both the identity authentication and access authorization issues in the distributed space network environment; 
\item Using virtual trust zone, the authentication mechanism ensures that only authenticated entities in same domain could communicate with each other, meanwhile the capability-based access control model provides a scalable, flexible, fine-grained and lightweight access scheme for space applications; 
\item A complete architecture of a blockchain-enabled decentralized access control system is properly designed, which includes identity authentication, capability token management and access right validation. The data structures of identity certificate and capability token are explained. The identity authentication algorithms and access right verification process are discussed in detail; and
\item A concept-proof prototype based on smart contracts is implemented both on resource-constrained edge devices and more powerful devices, and deployed on a local private Ethereum blockchain network; to emulate satellites with collection sensors, a ground-based system network, and SATCOM; respectively. A comprehensive experimental study has been conducted that evaluates the computational and the timeliness performance of using the public blockchain. 
\end{enumerate}

The remainder of this paper is organized as follows: Section \ref{sec:related} gives a brief review on the state-of-the-art research in access control and blockchain technology. Section \ref{sec:blendcac} illustrates the details of the proposed BlendCAC system and Section \ref{sec:prototype} explains the implementation of the proof-of-concept prototype. The experimental results and evaluation are presented in Section \ref{sec:experiment}. Finally, the summary, current limitations and on-going efforts are discussed in Section \ref{sec:conclusion}.

\section{Background Knowledge and Related Work}
\label{sec:related}

\subsection{SSA and DDDAS}
Space Situation Awareness (SSA) is considered as an important frontier because of the increased congestion of satellites vital for strategic decisions, communications, and weather/terrestrial observations. \cite{shen2017orbital}. Since the number of satellites in
orbit continues to grow exponentially, it is required to ensure that all spacecraft on-orbit work as intended to successfully accomplish their missions. 
The SSA environment generally consists of two major areas: satellite operations and space weather. The satellite operations are focused on the local perspective to enable continuous operations by understanding the space environment and build models to support satellite health monitoring (SHM). SSA is a systems design which utilizes data collected from ground and other space assets for RSO tracking, imaging and collision avoidance \cite{blasch2018dddas,chen2009comparison,faber2015randomized}. The key components of SSA include RSO tracking and characterization, satellite health monitoring and communication, information management, sensing, navigation, and data visualization \cite{blasch2018dddas,blasch2013enhanced}. 
To address satellite communications (SATCOM) challenge that requires cognitive spectrum management and agile waveform adaptation solutions, a game theory enabled high-level anti-interference strategy was proposed to solve interference in congested space environment \cite{shen2014network}. To support space defense analysis and mission trade-off investigations, a satellite orbital testbed (SOT) for space sensor resource allocation, is developed and evaluated for Pursuit-Evasion Game Theoretic Sensor Management \cite{shen2017orbital}.

DDDAS is a conceptual framework that synergistically combines models and data in order to facilitate the analysis and prediction of physical phenomena \cite{blasch2018dddas,dddas18}. In a SSA application, DDDAS is a variation of adaptive state estimation that uses computational feedback rather than physical feedback to enhance the information content of measurements. The feedback loops in DDDAS include a data assimilation loop and a sensor reconfiguration loop. The data assimilation loop calculates the physical system simulation by using sensor data error to ensure that the trajectory of the simulation more closely follows the trajectory of the physical system. As a fundamental aspect of DDDAS, the sensor reconfiguration loop seeks to manage the physical sensors in order to enhance the information content of the collected data. The simulation based on computational feedback process guides the sensor reconfiguration and the data collection, and in turn, improves the accuracy of the physical system environmental assessment (e.g., space weather and RSO tracking).  For sensor management, DDDAS develops runtime software methods for real-time control such as access control.

\subsection{Access Control Mechanism}
An Access Control (AC) mechanism, which specifies admittance to certain resources or services, contributes to the protection, security, and privacy for IT systems. As a fundamental mechanism to enable security in computer systems, AC is the process that decides who is authorized to have what communication rights on which objects with respect to some security models and policies \cite{gong1989secure}. An effective AC system is designed to satisfy the main security requirements, such as confidentiality, integrity and availability. In general, a comprehensive AC system addresses three main security issues: Authentication, Authorization and Accountability \cite{ouaddah2017access}. Authentication is the method of validating identity  based on registered entity's information. Authorization involves the following phases: defining a security policy (set of rules), selecting an AC model to encapsulate the defined policy, implementing the model, and enforcing the access rules \cite{ouaddah2017access}. Accountability employs audit logs to associate subjects with functions. 

The AC mechanism incorporates two aspects: the AC model and architecture. The Role-Based Access Control (RBAC) \cite{sandhu1996role} model provides a framework that authorizes user’s access to resources based on functions. In a RBAC model, permissions are assigned to each agent's role according to organizational functionality definition, and access rights are indirectly granted by associating a user with certain specified role. The functional role acts as an intermediary to bring users and permissions together. The RBAC model supports principles, such as least privilege, partition of administrative functions and separation of duties; and has been widely used in computer systems \cite{samarati2000access}. For example, the RBAC model implemented in IoT networks adopts a Web of Things (WoTs) framework \cite{de2008socrades,spiess2009soa}, which is a service-oriented approach, to enforce AC policies on the smart things via a web service application. However, current RBAC models are not able to address the key issues of implementing RBAC in a highly distributed network environment, such as self-management to handle the explosion of roles in complex and ambiguous space scenarios and autonomy to support physical objects through device-to-device communication.

Compared to the RBAC model, the Attribute-Based Access Control (ABAC) \cite{yuan2005attributed,smari2014extended}, model defines permissions based on any security relevant characteristics, known as attributes. In the ABAC, AC policies are defined by directly associating predefined attributes with subjects, resources, and conditions, respectively. Given all the attributes assignments, a Policy Rule, which is a Boolean function, decides whether to grant the subject's access to the resource under specific conditions. The ABAC model eliminates the definition and management of static roles used in the RBAC model. Hence, ABAC also eliminates the need for the administrative tasks for user-to-role assignment and permission-to-role assignment \cite{yuan2005attributed}. To address the weaknesses of the RBAC model in a highly distributed network environment, an ABAC extension to the AWS-IoTAC model was proposed to enhance the flexibility of AC in cloud-enabled IoT platform \cite{bhatt2017access}. An efficient Elliptic Curve Cryptography (ECC)-based authentication and the ABAC policy together was proposed to solve the resource-constrained problem of a perception layer \cite{ye2014efficient}. Although the ABAC is more manageable and scalable than the RBAC by providing finer-grained AC policies that involve multiple subject and object attributes, specifying a consistent definition of the attributes within a single domain or across multiple domains could significantly increase the effort and complexity of policy management as the number of devices grow, and a user-driven and delegation strategies are not supported with the ABAC model. Hence, the attribute-based proposal is not suitable for large-scale distributed network scenarios.

Capability-based Access Control (CapAC) utilizes the concept of capability that contains rights granted to the entity holding it \cite{ouaddah2017access}. The capability is defined as tokens, tickets, or keys that give the possessor permission to access an entity or object in a computer system \cite{dennis1966programming}. The CapAC has been implemented in many large-scale projects, like IoT@Work \cite{gusmeroli2012iot}. However, the direct application of the original concept of CapAC model in a distributed network environment has raised several issues, like capability propagation and revocation. To tackle these challenges, a Secure Identity-Based Capability (SICAP) System \cite{gong1989secure} was proposed to provide a prospective capability-based AC mechanism in distributed networks. By using an exception list, the SICAP enables monitoring, mediating, and recording capability propagations to enforce security policies as well as achieving rapid revocation capability \cite{gong1989secure}. By introducing a delegation mechanism for the capability generation and propagation process, a Capability-based Context-Aware Access Control (CCAAC) model was proposed to enable contextual awareness in federated devices \cite{anggorojati2012capability}. A federated delegation mechanism enables the CCAAC model has great advantages in terms of addressing scalability and heterogeneity issues in IoT networks. As a user-driven AC model, the CapAC supports machine-to-machine (M2M) communication, and presents great scalability and flexibility to deal with spontaneous and dynamic changes in distributed network environment. However, management of capability propagation becomes difficult without a proper delegation and revocation mechanism.

At the architecture level, the AC solutions are categorized as either the centralized or the decentralized approach. In a centralized AC architecture, the key components, like authorization policy management and policy decision making, employ a centralized authority. Outsourcing computational intensive tasks to a back-end cloud or a gateway relieves smart device from the burden of handling AC related functionalities. The approaches in \cite{zhang2010extended,bhatt2017access,ye2014efficient,anggorojati2012capability} are centralized methods. The centralized AC solutions present many advantages, such as easy to adapt existing AC model and simple security policy management. However, enforcing AC on the centralized architecture also suffers from several data management drawbacks, like single point of failure, performance bottleneck and privacy issues.

To address the drawbacks in the centralized AC solutions, designing a decentralized AC mechanism supports SSA performance. A distributed Capability-based Access Control (DCapAC) mechanism was proposed that directly deploy AC on resource-constrained devices \cite{hernandez2013distributed}. The DCapAC allows smart devices to autonomously make decisions on access rights based on an authorization policy, and it has advantages in scalability and interoperability. To address challenges like scalability, granularity, and dynamicity in AC strategies for SSA systems; a Federated Capability-based Access Control model (FedCAC) is proposed to enable an effective AC mechanism to devices, services and information in large scale SSA systems \cite{xu2018federated}. Migrating part of the centralized authorization validation tasks to local devices helps the FedCAC to be lighter and context-awareness enabled. The decentralized AC approach presents advantages, like supporting User-driven security mechanism and not relying on centralized trust authority. Nonetheless, the decentralized approach also brings many issues, such as requiring a complex AC mechanism and introducing overhead such as in satellite communications.

\subsection{Blockchain and Smart Contract}
The blockchain technology, which was initially introduced by Nakamoto in 2008 \cite{nakamoto2008bitcoin}, has demonstrated its success in decentralization of digital currency and payment, like bitcoin. A Blockchain is a replicated public database (Ledger) that records, stores, and updates all data as chained blocks. It is a public ledger that provides a verifiable, append-only chained data structure of transactions. Enforcing the consensus mechanism on a peer-to-peer network framework, the blockchain is essentially a decentralized architecture that does not rely on a centralized authority. The transactions are approved by a large amount of distributed nodes called miners and recorded in timestamped blocks, where each block is identified by a cryptographic hash and chained to preceding blocks in a chronological order. Blockchain uses a consensus mechanism, which is enforced on miners, to maintain the sanctity of the data recorded on the blocks. Thanks to the trustless proof mechanism running on miners across networks, users can trust the system of the public ledger stored worldwide on many different nodes maintained by "miner-accountants", as opposed to having to establish and maintain trust with a transaction counter-party or a third-party intermediary \cite{swan2015blockchain}. Thus, Blockchain is considered an ideal decentralized architecture to ensure distributed transactions between all participants in a trustless environment.

Emerging from the smart property, a \emph{smart contract} allows users to achieve agreements among parties and supports a variety of flexible transaction types through blockchain networks. By using cryptographic and security mechanisms, smart contract combines protocols with user interfaces to formalize and secure relationships over computer networks \cite{szabo1997formalizing}. A smart contract includes a collection of pre-defined instructions and data that have been saved at a specific address of a blockchain as a Merkle hash tree, which is a constructed bottom-to-up binary tree data structure. Since smart contracts are developed as scripts and stored on the blockchain, each smart contract has a unique address that resides on the blockchain. Through exposing public functions or application binary interfaces (ABIs), a smart contract interacts with users to offer a predefined business logic or contract agreement. 

Through encapsulating operational logic as a bytecode and performing Turing complete computation on distributed miners, a smart contract allows the user to trans-code more complex business models as new types of transactions on a blockchain network. A smart contract provides a promising solution to allow the implementation of more flexible and complex applications far beyond cryptocurrencies, such as data provenance, resource sharing and dynamic spectrum access, etc. The blockchain and smart contract enabled security mechanism for applications has been a hot research topic and some efforts have been reported recently, for example, smart surveillance system \cite{nagothu2018microservice,nikouei2018real}, social credit system \cite{xu2018constructing}, identity authentication \cite{hammi2018bubbles} and access control \cite{xu2018smartcac,xu2018blendcac}. 
The research report by R.Mital \cite{mital2018blockchain} presents the potential role, capabilities and value of blockchain and smart contract usage within constellation and swarm satellite architectures. The blockchain and smart contract together are promising towards providing a decentralized solution to allow more flexible and fine-grained AC models on space applications.

\section{BlendCAC: A BLockchain-ENabled Decentralized CapAC Mechanism}
\label{sec:blendcac}
\begin{figure}
\begin{center}
\begin{tabular}{c}
\includegraphics[height=9.8cm]{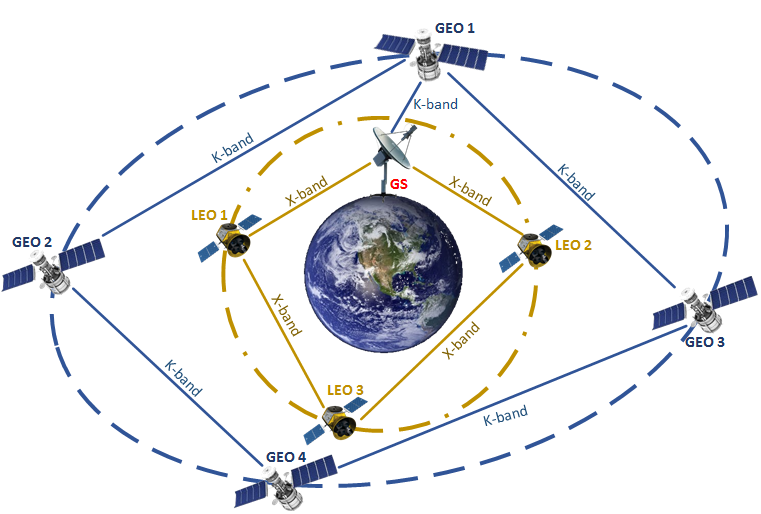}
\end{tabular}
\end{center}
\caption 
{ \label{fig0:ssa} Research Scenarios of SSA. } 
\end{figure}

To support sensors and systems for SSA applications, there is a need for resilient, reliable, and robust designs. The space system comprises sensors to monitor the environment, communications to transfer information, and algorithms to process the data such as information fusion \cite{zheng2018multispectral}. Examples of functions include: evaluating health of the system, tracking of space objects both on-orbit optical observations and ground observations, as well as passing messages with secure communication. Figure \ref{fig0:ssa} demonstrates a sketch of the research scenarios of SSA. There are four geostationary (GEO) satellites (GEO 1, GEO 2, GEO 3, and GEO 4) for the blue orbit and three Low Earth Orbit (LEO 1, LEO 2, and LEO 3) satellites for the yellow orbit. One Ground Site (GS) is used for ground observations, which provides spectrometers analyzing optical emissions from space object thrusters. Optical observations are performed on satellites by using an on-board camera to confirm plume emission and take images of the thruster in operation. While ground observations are conducted on ground sites to determine the emission spectra of actively firing Hall thrusters in vacuum chambers, the data transmission among satellites and the ground sites is carried out on different satellite communications (SATCOM) channels.

As Fig. \ref{fig0:ssa} demonstrates, LEO uses the X-band SATCOM communication channel (yellow lines), while GEO utilizes the K-band SATCOM communication channel (blue lines). The collected data from both on-orbit optical observations and ground observations is shared to provide services for SSA applications. Data integrity and ccess security is significant to ensure data integrity in SSA applications. Thus, a flexible and fine-grained access control scheme is necessary to ensure that data sharing among authenticated space objects and cooperative operations are performed by authorized entities. Furthermore, the satellite communications (SATCOM) infrastructure includes heterogeneous satellite communication technologies, hybrid space-terrestrial systems, and a decentralized access control architecture.

\subsection{BlendCAC System Architecture for SSA}
\begin{figure}
\begin{center}
\begin{tabular}{c}
\includegraphics[height=10.5cm]{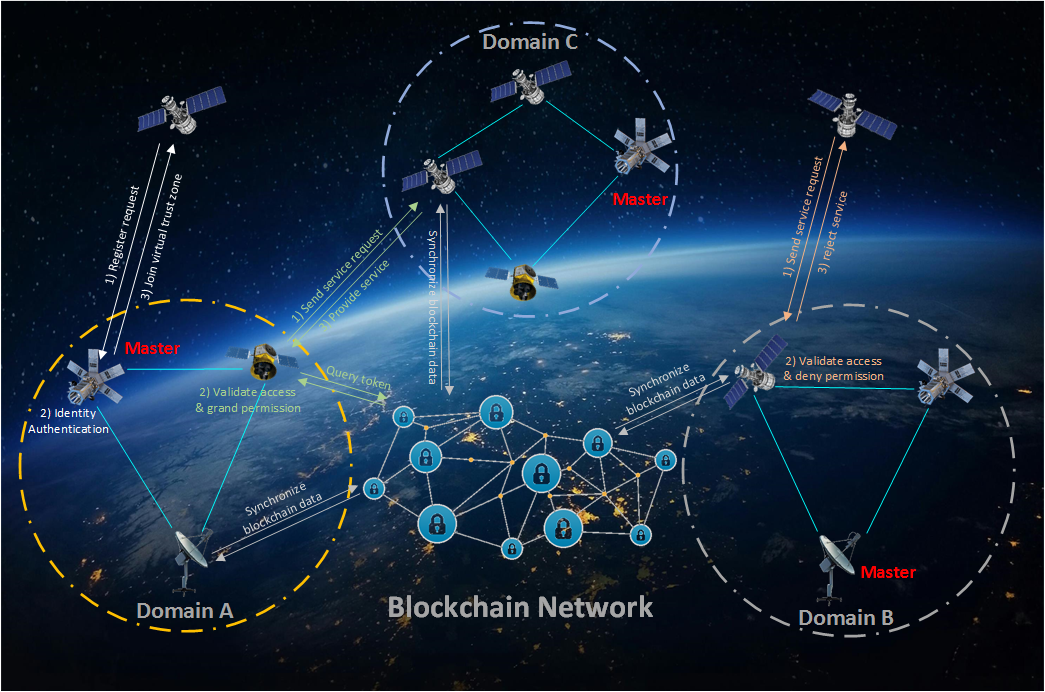}
\end{tabular}
\end{center}
\caption 
{ \label{fig1:arch} System Architecture of BlendCAC. } 
\end{figure}

Inspired by the smart contract and blockchain technology, a decentralized federated capability-based AC framework for SSA systems, called BlendCAC, is introduced in this paper, and a prototype of proposal is implemented in a physical network environment to verify the efficiency and effectiveness on a simulated space network scenario.  Figure \ref{fig1:arch} illustrates the proposed BlendCAC system architecture for SSA, which is intended to function in a scenario including multiple isolated space service domains without pre-establishing a trust relationship. 

In the proposed access control framework shown in Fig. \ref{fig1:arch}, each domain master works as a certificate authority to provide identity authentication services, and enforces delegated security policies to manage domain related space object functions and services. Both satellites and ground sites could become domain masters. 
The identification authentication and access authorization policies are transcoded to the smart contracts and deployed across the blockchain network, and identity validation, authorization delegation and access right verification are developed as service applications running on both domain masters and resident space objects in the space network. The operation and communication modes are listed as follows:

\begin{enumerate}
\item \emph{Identification Authentication}: All entities must create at least one main account defined by a pair of keys to join the blockchain network. Each account is uniquely indexed by its address that is derived from his/her own public key. Thus, account address is ideal for identity authentication in the BlendCAC system given assumption that authentication process is ensured by a blockchain network. In this scenario, a virtual trust zone is created by the domain master, such that each object is allowed to communicate with objects in the same virtual trust zone. Entity registration process uses account address as a Virtual Identity (VID), which are recored in the profile database that is deployed on the domain master. A new entity must send authentication request to the domain master in order to join the virtual trust zone. Once the identity information related to requester is verified, the domain master will create the ticket for the registered entity by recording his/her blockchain account address and group ID in the blockchain for authentication process when an service request happens. As a result, the domain masters not only are responsible for identification authentication, but also are able to enforce delegated authorization policies and perform decision-making to directly control the objects or resources in virtual trust zone instead of depending on third parties.  

\item \emph{Smart Contract Deployment}: The smart contract, which carries out authentication and manages federated delegation relation and capability tokens, must be developed and deployed on the blockchain network by the policy owner. In the BlendCAC framework, the administrators of the DDDAS, who act as the data and policy owners, could deploy smart contracts encapsulating authentication and authorization algorithms. After smart contracts have been deployed successfully on the blockchain network, they become visible to the entire network owing to the transparency and publicity properties of the blockchain protocol, which means that all participants in the blockchain network can access transactions and smart contracts recorded in the chain data. Thanks to the cryptographic and security mechanisms provided by the blockchain network, smart contracts can secure any algorithmically specifiable protocols and relationships from malicious interference by third parties under a trustless network environment. After synchronizing the blochchain data, all nodes could access all transactions and recent state of each smart contract by referring local chain data. Each node interacts with the smart contract through the provided contract address and the Remote Procedure Call (RPC) interface.  

\item \emph{Capability Authorization}: To successfully access services or resources at service providers, an entity initially sends an access right request to the domain master to get a capability token. Given the entity's ticket, which is the authenticated identity information saved in the blockchain, a decision making policy module running on the domain master evaluates the access request by enforcing the authorization policies. If the access request is granted, the domain master issues the capability token encoding the access right, and then launches a transaction to update the token data in the smart contract. Once the transaction has been approved and recorded in a new block, the domain master responds to the requester by providing a smart contract address for the querying token data. Otherwise, the access right request is rejected by returning denied acknowledgement. 

\item \emph{Access Right Validation}: 
The authorization validation process is performed at the space object who works as the local service provider to receive space service requests from subjects, such as querying satellite sensors of observed spectrometer optical emissions and images of the thruster operations.
Through regularly synchronizing the local chain data with the blockchain network, a service provider just simply checks the current state of the contract in the local chain to get a capability token associated with the entity's address. Considering the capability token validation and access authorization process result, if the access right policies and conditional constraints are satisfied, the service provider grants the access request and offers services to the requester. Otherwise, the service request is denied.
\end{enumerate}

To enable a scalable, distributed and fine-grained AC scheme for space networks, the proposed BlendCAC is focused on three issues: the identification authentication, the identity-based capability management, and the access right authorization.

\subsection{Identification Authentication}
Authentication is the mechanism of validating identity information of entities. The overall purpose of an authentication strategy is to allow multiple entities to communicate with each other in a trustworthy way in a trustless network environment. Inspired by the idea of bubble of trust, all members in a bubble zone can trust each other \cite{hammi2018bubbles}. The scheme in Figure \ref{fig2:auth} illustrates the proposed authentication approach and all the phases of the ecosystem's life-cycle. The involved phases in an authentication process is as follows:

\begin{figure}
\begin{center}
\begin{tabular}{c}
\includegraphics[height=13.2cm]{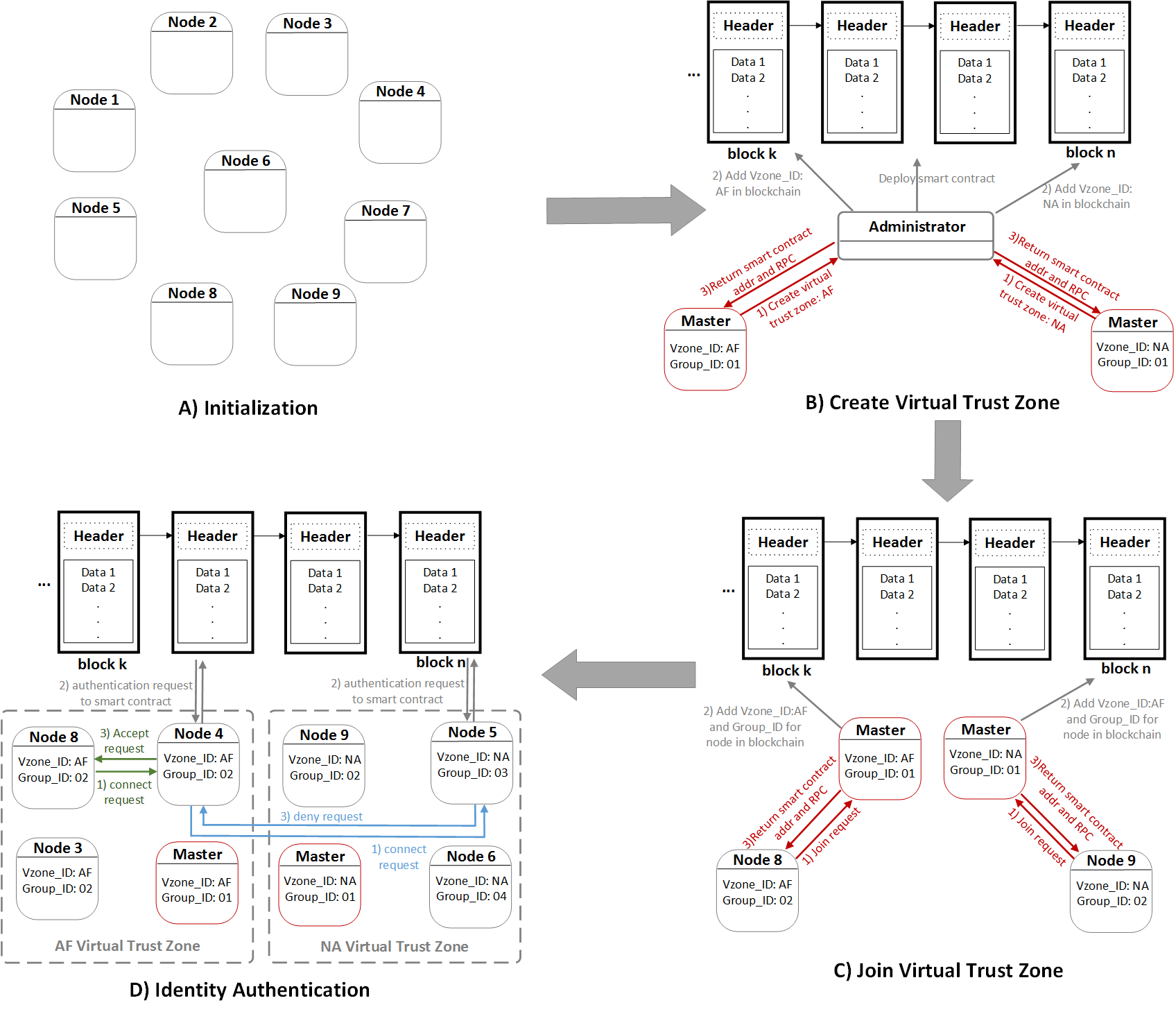}
\end{tabular}
\end{center}
\caption 
{ \label{fig2:auth} Virtual Trust Zone Mechanism for Authentication. } 
\end{figure} 

\begin{itemize}
\item \emph{Initialization}: As shown in Fig. \ref{fig2:auth}A, 
the connected space objects that belong to different areas freely communicate with each other. All the space objects in the network could be categorized as either domain masters or service nodes. All the entities should use their main blockchain account addresses as the virtual ID for authentication and access control purposes. 

\item \emph{Creation of Virtual Zones}: The smart contract for authentication is created by administrators, and deployed through transaction, as shown in Figure \ref{fig2:auth}B. The master has to communicate with the administrator to create the virtual trust zone for their domain. The master sends a transaction that contains the master's identifier as well as the identifier of the group to be created. The administrator verifies the identity information of the master and sends the transaction to the smart contract to create the virtual trust zone for  the master. The blockchain verifies the uniqueness of both of the group ID and the master's Virtual ID. If all conditions are satisfied, the smart contract generates a new virtual trust zone with an unique virtual zone ID for the master and returns the smart contract address and authorized RPC for the master to interact with.

\item \emph{Join Virtual Trust Zone}: Figure \ref{fig2:auth}C demonstrates how nodes are associated with the virtual trust zone. After a virtual trust zone has been created, the nodes in turn, send transactions to the master in order to join their respective virtual trust zones. The domain master checks the applicant's identifier based on the registration policy. If all conditions are satisfied and the applicant has never joined the zone before, the master interacts with the smart contract to add the node to the virtual trust zone. As miners have verified the transaction and generated a new block, the node joins the virtual trust zone successfully with an unique group ID.

\item \emph{Identity Authentication}: As all nodes' group information are recored in the blockchain and the identifier verification process is ensured by the smart contract, the identity authentication is no longer relying on a third-party centralized trust authority. In Figure \ref{fig2:auth}D, node 8 could successfully talk to node 4 owning to the fact that they have the same Vzone\_ID. However, the node 5 denies the connect request from node 4 because they do not share the same Vzone\_ID, which means that they actually do not belong to the same virtual trust zone.
\end{itemize}

Through clustering the nodes into different virtual trust zones, the application domains become isolated. Only those authenticated entities are allowed to communicate with group members of their zones, while any entities outside of the zones are considered as suspicious and prevented from being connected to any group members in the zones.

\subsection{Capability Token Structure}
In the access authorization scenarios, the entities are categorized as subjects or objects. \textit{Subjects} are defined as entities who request services from the service providers, while \textit{objects} are referred to entities who offer the resources or services. Entities could be either human operators or resident space objects, like satellites. Since the identity registration and authentication processes are mainly performed on domain masters, a profile database that is used for recording profile of each group member is constructed and maintained by the domain master. In the profile databased, all registered entities are associated with a globally unique Virtual Identity (VID), which is used as the prime key for searching entities' profile information. As each entity has at least one main account indexed by its address in the blockchain network, the blockchain account address is used to represent the VID for profiling register entities.

In general, the capability specifies which subject can access resources at a target object by associating subject, object, actions and condition constraints. The \emph{identity-based capability} structure is defined as a hash table which is represented as follows:

\begin{equation}
\label{eq:ICap}
ICap = f(VID_{S}) {\rightarrow} \{VID_{O}, OP, C\}
\end{equation}

\noindent{where the parameters are:}
\begin{itemize}
\item $f$: a one-way hash mapping function to establish relation between a subject and authorized internal capability set; 
\item $VID_{S}$: the virtual ID of a subject that requests an access to a service
or resource;
\item $VID_{O}$: the virtual ID of an object that provides a service or resource;
\item $OP$: a set of authorized operations, e.g. read, write, execute; and
\item $C$: a set of context awareness information, such as time, location, etc.
\end{itemize}

In the AC system, the elements in $OP$ set can be represented as actions, such as $\{Read\}$, $\{Write\}$, $\{Read; Write\}$, or $\{NULL\}$. If $OP=\{NULL\}$, any operation conducted on the resource is not allowed. $C$ is defined as a context constraints set, like $C=\{C1, C2\}$ or $C=\{NULL\}$. If $C=\{NULL\}$, no context constraint is considered in the access right validation process.

\subsection{Capability-based Access Right Authorization}

\begin{figure} [t]
\begin{center}
\begin{tabular}{c}
\includegraphics[height=7.5cm]{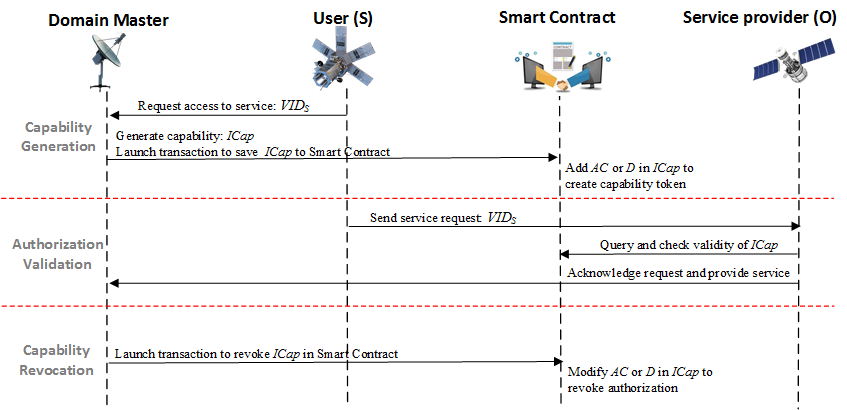}
\end{tabular}
\end{center}
\caption[example] { \label{fig:3-CapAuthorization} Flowchart of the Capability-based Access Right Authorization.}
\vspace{-10pt}
\end{figure}

The capability token structure and the related operations are transcoded to a smart contract that is deployed on the blockchain network, while the access right (AR) authorization is implemented as a policy-based decision making service running on the domain master. As shown by Figure \ref{fig:3-CapAuthorization}, a comprehensive capability-based access right authorization procedure consists of four steps: capability generation, access right validation, capability delegation and revocation.

\begin{enumerate}
\item \emph{Capability Generation}: As one type of meta data to represent the access right, the capability $ICap$ could be generated by associating a VID with an AR, thus the $ICap$ has the identified property to prevent forgery. After receiving an access request from a user, the domain master generates a capability token based on the access right authorization policy, and launches transactions to save a new token data to a smart contract. A large number of $ICap$'s are grouped into the capability pools on the smart contract, which could be proofed and synchronized among the nodes across the blockchain network.

\item \emph{Access Right Validation}: After receiving the service request from a subject, the service provider first fetches the capability token from the smart contract using the subject's address, then makes decisions on whether or not to grant an access to the service according to the local AC policy. Implementing access right validation at the local service provider allows smart objects to be involved in the AC decision making task, which is suitable to offer a flexible and fine-grained AC service in distributed space networks.

\item \emph{Capability Revocation}: The capability revocation considers two scenarios: partial access right revocation and $ICap$ revocation. In the system design, only the administrator or domain masters are allowed to perform revocation operation on the capability tokenized smart contract. In the partial access right revocation process, the authorized entities could remove part of the entries from $AR$ to revoke the selected access rights. In case of $ICap$ revocation, through directly clearing the $AR$ in $ICap$, the whole capability token becomes unavailable to all associated entities.
\end{enumerate}

\section{Prototype Design}
\label{sec:prototype}
A proof-of-concept prototype system has been implemented in a real private Ethereum blockchain network environment  extending an SSA study using a cloud architecture \cite{liu2014adaptive}. As the second biggest ledger in the world, Ethereum is robust against attacks and data falsifications. In addition, transactions in Ethereum adopt the Elliptic Curves Cryptography as the signature scheme, which represents robust and lightweight properties for constrained devices. Furthermore, compared with other open blockchain platforms, like Bitcoin and Hyperledger, Ethereum has a more matured ecosystem and is designed to be more adaptable and flexible for the development of a smart contract and business logic \cite{ehtereum}. 

\subsection{Authentication Certificate and Capability Token Structure}
The proposed identity authentication and AC models have been transcoded to smart contracts using Solidity \cite{solidity}, which is a contract-oriented, high-level language for implementing smart contracts. With Truffle \cite{truffle}, which is a world class development environment, testing framework and asset pipeline for Ethereum, contract source codes are compiled to Ethereum Virtual Machine (EVM) bytecode and migrated to the Ethereum blockchain network.

To implement a BlendCAC system on resident space objects without introducing significant overhead over 
SATCOM communication and computation,
delegation certificate and capability token data structure is represented in JSON \cite{crockfordrfc} format. Compared to XML-based language for AC, like XACML and SAML, JSON is lightweight and suitable for resource constrained platforms.

\begin{figure} [ht]
\begin{center}
\begin{tabular}{c}
\includegraphics[height=12.5cm]{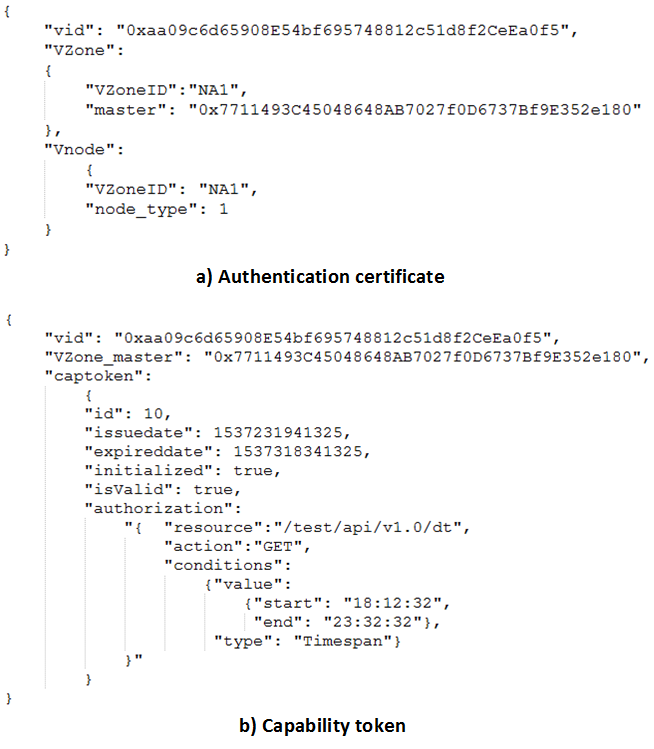}
\end{tabular}
\end{center}
\caption[example] { \label{fig:4-token_data} Token Data Structure used in BlendCAC.}
\end{figure}

Figure \ref{fig:4-token_data}(a) demonstrates an example of the authentication certificate, and the data fields in the data structure are described as follows:

\begin{itemize}
\item $vid$ : a 20-byte value to represent address of the certificate owner in the blockchain network;
\item $VZone$: a virtual trust zone data that has been created by the master, including
	\begin{itemize}
	\item $VZoneID$: a string that is used for a virtual trust zone data uniquely represented; and
	\item $master$: a 20-byte value used to represent the blockchain account address of the entity who created the virtual trust zone.
	\end{itemize}
\item $Vnode$: a set of identity information that has been associated to the node for authentication, including
	\begin{itemize}
	\item $VZoneID$: a string that is used to record the unique ID of a virtual trust zone, in which the entity has participated; and
	\item $node\_type$: an integer to specify the role that the entity has been assigned in the virtual trust zone, either the master or the follower.
	\end{itemize}
\end{itemize}

Figure \ref{fig:4-token_data}(b) presents an example of the capability token data used in the AC mechanism. A brief description of each field is provided as follows:

\begin{itemize}
\item $vid$ : a 20-byte value to represent address of the capability token owner in the blockchain network;
\item $VZone\_master$ : a 20-byte value used to record address of the master in the virtual trust zone that entity has joined;
\item $id$: the auto-incremented prime key to identify a capability token;
\item $initialized$: a bool flag used for checking token initialized status;
\item $isValid$: a boolean flag signifying the enabled status to show whether or not the token is valid;
\item $issuedate$: for identifying the date and time when the token was issued;
\item $expireddate$: the date and time when a token expires;
\item $authorization$: a set of access right rules that the issuer has granted to the subject, including
	\begin{itemize}
	\item $action$: to identify a specific granted operation over the resource;
	\item $resource$: to grant the operation in the service provider; in this case, the resource is defined as the granted REST-ful API; and
	\item $conditions$: a set of conditions which must be fulfilled locally on the service provider to grant the corresponding operation.
	\end{itemize}
\end{itemize}

After a smart contract has been successfully deployed on the blockchain network, all nodes in the network could interact with the smart contract using address of the contract and the Application Binary Interface (ABI) definition, which describes the available functions of the contract.

\subsection{Identity Authentication Policy Service}

The identity authentication mechanism based on the virtual trust zone is implemented as a set of service interface functions, which are executed by the smart contract to enforce the authentication policy. \textit{Algorithm 1} illustrates the virtual trust zone construction process. The function $createVZone()$ receives the inputs of the string $VZoneID$, and returns the $VZone$ creation result. The process firstly checks the entity address so that the supervisor or the valid masters are allowed to create a new VZone. The existing virtual trust zones could be deleted either by the supervisor or masters who have created the VZones, and the revocation process is illustrated by \textit{Algorithm 2}.

After the virtual trust zones have been constructed by masters, other nodes could send registration requests to the masters for joining the virtual trust zone. \textit{Algorithm 3} describes the process that how a node becomes a member of a VZone. Once the Vnode of the applicant has been recorded in the blockchain, he/she could communicate with other nodes in the same VZone. The associated trust relationship between a node and the VZone could be revoked either through leave request sent by the node, or directly be removed by the supervisor and the master of the VZone. \textit{Algorithm 4} explains the operation to remove a node from a virtual trust zone.

The identity verification is enforced based on service providing scenarios. As a service provider received a service request from an entity, he/she just queries entity's Vnode data in the blockchain and verifies the identification by simply checking whether or not the entity has the same VZoneID. The requests from non-VZone entities are directly rejected.

\begin{algorithm}\linespread{1.1}
\caption{Create Virtual Trust Zone}
\small 
\begin{algorithmic}[1]
\REQUIRE $VZoneID$
\STATE $entityAddr = msg.sender$
\IF{$(entityAddr==supervisor)$  or  $(isValidMaster(entityAddr) == true)$}
	\IF{$if(Vzone[VZoneID].master == address(0)) $}
		\STATE $Vzone[VZoneID].uid += 1$
        \STATE $Vzone[VZoneID].master = entityAddr$
		\STATE $Vnode[entityAddr].VZoneID = VzoneID$
        \STATE $Vnode[entityAddr].node\_type = 1$
        \STATE $return  True$
	\ELSE
		\STATE $return  False$
    \ENDIF
\ELSE
	\STATE $return False$
\ENDIF
\end{algorithmic}
\end{algorithm}

\begin{algorithm}\linespread{1.1}
\caption{Revoke Virtual Trust Zone}
\small 
\begin{algorithmic}[1]
\REQUIRE $VZoneID$
\STATE $entityAddr = msg.sender$
\IF{$entityAddr==supervisor$}
\IF{$(entityAddr==supervisor)$  or  
$((isValidMaster(entityAddr) == true)$ and $Vzone[VZoneID].master == entityAddr)$}
		\STATE $curr\_master = Vzone[VzoneID].master$
		\STATE $Vzone[VZoneID].uid += 1$
        \STATE $Vzone[VZoneID].master = address(0)$
		\STATE $Vnode[entityAddr].VZoneID = ""$
        \STATE $Vnode[entityAddr].node\_type = 0$
        \STATE $return  True$
	\ELSE
		\STATE $return  False$
    \ENDIF
\ELSE
	\STATE $return False$
\ENDIF
\end{algorithmic}
\end{algorithm}

\begin{algorithm}\linespread{1.1}
\caption{Join Virtual Trust Zone}
\small 
\begin{algorithmic}[1]
\REQUIRE $VZoneID$
\REQUIRE $nodeAddr$
\STATE $entityAddr = msg.sender$
\IF{$(entityAddr==supervisor)$  or  $(entityAddr == Vzone[VZoneID].master)$}
	\IF{$Vnode[nodeAddr].node_type == 0$}
		\STATE $Vnode[nodeAddr].VZoneID = VzoneID$
        \STATE $Vnode[nodeAddr].node\_type = 2$
        \STATE $return  True$
	\ELSE
		\STATE $return  False$
    \ENDIF
\ELSE
	\STATE $return False$
\ENDIF
\end{algorithmic}
\end{algorithm}

\begin{algorithm}\linespread{1.1}
\caption{Leave Virtual Trust Zone}
\small 
\begin{algorithmic}[1]
\REQUIRE $VZoneID$
\REQUIRE $nodeAddr$
\STATE $entityAddr = msg.sender$
\IF{$(entityAddr==supervisor)$  or  $(entityAddr == Vzone[VZoneID].master)$}
	\IF{$Vnode[nodeAddr].node_type == 2$}
		\STATE $Vnode[entityAddr].VZoneID = ""$
        \STATE $Vnode[entityAddr].node\_type = 0$
        \STATE $return  True$
	\ELSE
		\STATE $return  False$
    \ENDIF
\ELSE
	\STATE $return False$
\ENDIF
\end{algorithmic}
\end{algorithm}

\subsection{Access Authorization Service}
The access authorization and validation policy is enforced as a web service application, which emulates the space service scenarios among satellites and ground sites. The test application is developed on the Flask framework \cite{flask} using Python.
The Flask is a micro-framework for Python based on Werkzeug, Jinja 2 and good intentions. The lightweight and extensible micro architectures make the Flask a preferable web solution on resource constrained devices.

\begin{figure} [ht]
\begin{center}
\begin{tabular}{c}
\includegraphics[height=9cm]{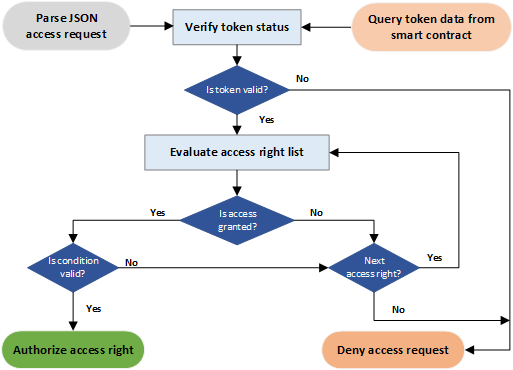}
\end{tabular}
\end{center}
\caption[example] { \label{fig:5-Cap_validation} Access Authorization Process of BlendCAC.}
\end{figure}

Web service application in the BlendCAC system consists of two parts: client and server. The client performs operations on resource by sending data request to the server, while the server provides REST-ful API for the client to obtain data or perform operations on the resource at the server side. A capability based AC scheme is enforced at the server side by performing access right validation on the service provider. The access right validation process is launched after a request containing the client's identity is received by the server. Figure \ref{fig:5-Cap_validation} shows a block diagram with the steps to process an authorization request.

\begin{enumerate}
\item \emph{Check cached token data}: After receiving a service request from a user, the service provider firstly checks whether or not the token data associated with user's address exists in the local database. If it is failed in searching the token data, the service provider can fetch the token data from the smart contract through calling an exposed contract method and save token data to the local database. Otherwise, the token data is directly reloaded from the local token database for further validation. The service provider regularly synchronizes the local database with the smart contract to ensure the token data consistency.

\item \emph{Verify token status}: As a capability token has been converted to the JSON data, the first step of token validation is checking the current capability token status, such as initialized, isValid, issuedate, and expireddate. If any status of a token is not valid, the authorization process stops and sends a deny access request acknowledgement back to the subject.

\item \emph{Check whether access is granted or not}: The service provider will go through all access rules in the access right set to guarantee that the request operation is permitted. The process checks whether or not the REST-ful method used by the requester matches the authorized action of current access rules and the value of the resource field is the same as the Request-URI option used by the requester. If the current access rule verification failed, the process skips to the next access rule for evaluation. If none of the access rules could successfully pass the verification, the authorization validation process stops and denies the access request.

\item \emph{Verify the conditions}: Even though the action on a target resource is permitted after the access validation, the context-awareness constraints are necessary to be evaluated on the local device by verifying whether or not the specified conditions in the token are satisfied. The condition verification process goes through all constraints in the condition set to find the matched ones. If no condition is fulfilled in the given local environment, the access right validation process stops and denies access request.
\end{enumerate}

\section{Experimental Study}
\label{sec:experiment}
In order to evaluate the performance and the overhead of our AC scheme, the identity authentication and access authorization are transcoded to separate smart contracts and enforced on the experimental web service system. The profiles and policy rules management are developed by using an embedded Structured Query Language (SQL) database engine, called SQLite \cite{sqlite}. The lower memory and computation cost make the SQLite an ideal database solution to resource constrained system, like Raspberry Pi. 
All documents and source code are available on BC\_DDDAS project repository on GitHub\cite{dddas}.

\subsection{Testbed Setup}
The mining task is performed on a system with stronger computing power, like a laptop or a desktop. Two miners are deployed on a laptop and other four miners are distributed on four desktops. Table \ref{tab:testbed} describes configuration of nodes used in the experiments. In our system, the laptop acts as a cloud computing server, while all desktops work as fog computing nodes to take role of domain masters. Each miner uses two CPU cores for mining. The edge computing services are deployed on two Raspberry PI 3 Model B. Since the Raspberry PI is not powerful enough to carry out mining task, so all Raspberry Pi devices function as nodes to participate the private blockchain network without mining. All devices use Go-Ethereum \cite{goethereum} as the client application to work on the blockchain network.

\begin{table}[ht]
\caption{Configuration of Experimental Nodes.} 
\label{tab:testbed}
\begin{center}       
\begin{tabular}{|l|p{3.7cm}|p{3.7cm}|p{4cm}|} 
\hline
\rule[-1ex]{0pt}{3.5ex} \textbf{Node Device} & Lenovo P50 & Dell Optiplex 760 & Raspberry Pi 3 Model B \\
\hline\hline
\rule[-1ex]{0pt}{3.5ex} \textbf{CPU} & 2.3 GHz Intel Core i7 (8 cores) & 3 GHz Intel Core TM (2 cores) & quad-core ARM Cortex A53, 1.2GHz \\
\hline
\rule[-1ex]{0pt}{3.5ex} \textbf{Memory} & 16GB DDR3 & 4GB DDR3 & 1GB SDRAM \\
\hline
\rule[-1ex]{0pt}{3.5ex} \textbf{Storage} & 250G SSD+ 500G HHD & 250G HHD & 32GB (microSD card) \\
\hline
\rule[-1ex]{0pt}{3.5ex} \textbf{Operation System} & Ubuntu 16.04 & Ubuntu 16.04 & Raspbian GNU/Linux 8 (jessie) \\
\hline
\end{tabular}
\end{center}
\end{table}

\subsection{Experimental Results}

\begin{figure} [t]
\begin{center}
\begin{tabular}{c}
\includegraphics[height=14 cm]{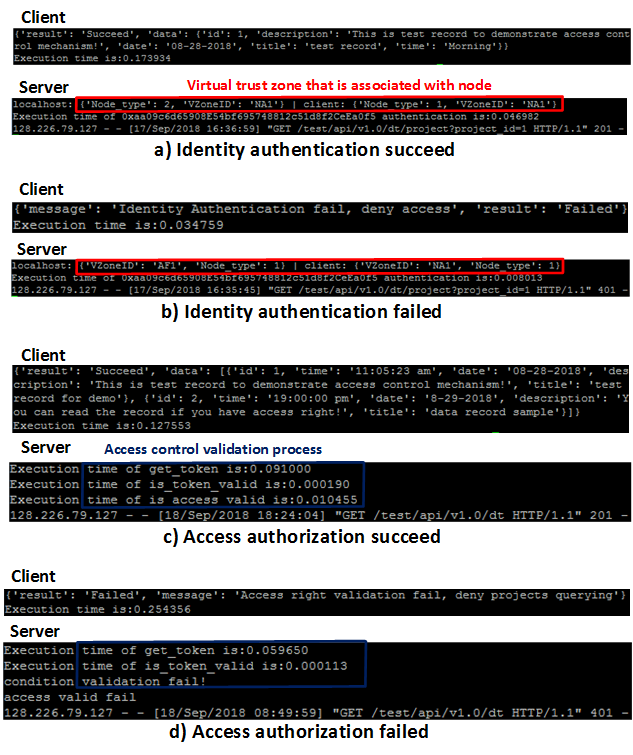}
\end{tabular}
\end{center}
\caption[example] { \label{fig:6-CapACResult} Examples of Experimental Results of the BlendCAC System.}
\vspace{-10pt}
\end{figure}

To verify the effectiveness of the BlendCAC approach against unauthorized access requests, a service access experiment is carried out on a simulated SATCOM network.
In the simulation environment, the edge devices represent satellites and the server is the ground communication receiving data, such as space imagery. In the test scenario, one Raspberry Pi 3 device works as the client and another works as the service provider. The identity authentication results are shown in Figs. \ref{fig:6-CapACResult} (a)-(b).  Figure \ref{fig:6-CapACResult} (a) demonstrates that the node '0xaa09c6d65908e54bf695748812c51d8f2ceea0f5' successfully passed the authentication process executed on the server with the same VZoneID. Figure \ref{fig:6-CapACResult} (b) shows a failed authentication scenario caused by communicating with entity who belongs to a different virtual trust zone.

Given the access authorization process shown in Figs. \ref{fig:6-CapACResult} (c)-(d), when any of the steps in the authorization procedure fails, the running process immediately aborts instead of continuing to step through all the authorization stages. As shown by Fig. \ref{fig:6-CapACResult}(d), the server stopped the authorization process due to the failure in verifying the granted actions or the conditional constraints that are specified in the access right list. Consequently, the client node received a deny access notification from the server and cannot read the requested data. In contrast, Fig. \ref{fig:6-CapACResult}(c) presents a successful imagery data request example, in which the whole authorization process is accomplished at the server side without any error. Finally, the client successfully retrieves the imagery data from the service provider.

\subsection{Performance Evaluation}
In the simulation environment, two Raspberry Pi devices emulate satellites to provide on-orbit observations while a desktop computer serves as a ground site to perform ground observations. one Raspberry Pi device is adopted to play the role of the client, while another Raspberry Pi and desktop computer are service providers, and an Ethernet is used to simulate SATCOM communication channel. 
To measure the general cost incurred by the proposed BlendCAC scheme both on the space (e.g., satellite) devices' processing time and the network communication delay, 100 test runs have been conducted based on the proposed test scenario, where the client sends a data query request to the server for an access permission. This test scenario is based on an assumption that the subject has joined the virtual trust zone and has been assigned a valid capability token when it performs the action. Therefore, all steps of identity authentication and authorization validation must be processed on the server side so that the maximum latency value is computed.

\begin{figure} [t]
\begin{center}
\begin{tabular}{c} 
\includegraphics[height=8.5cm]{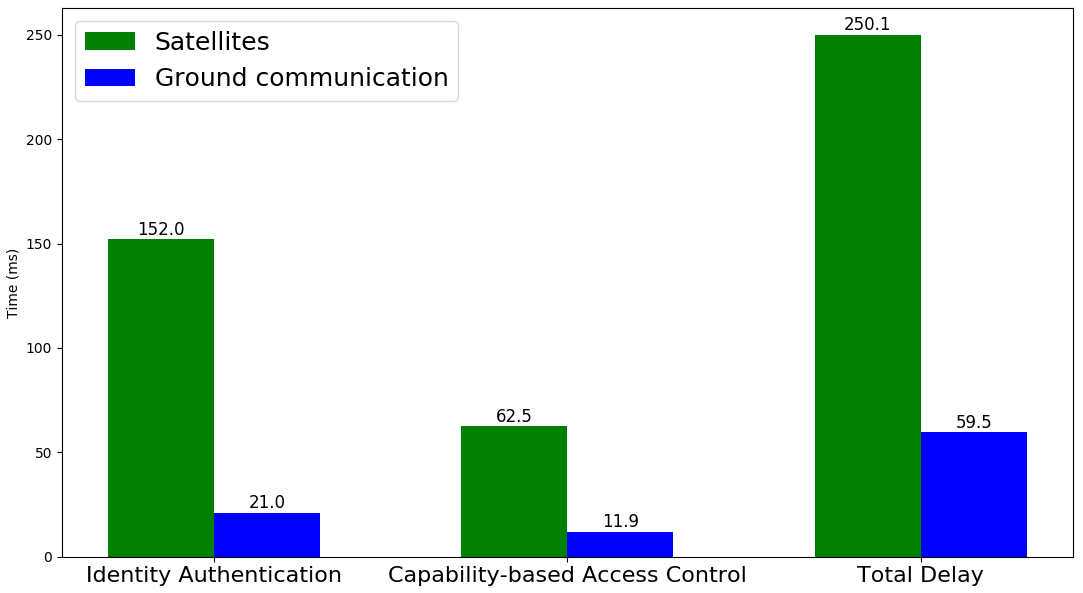}
\end{tabular}
\end{center}
\caption[example] { \label{fig:7-CapAc_exec_time} Computation Time for Each Stage in BlendCAC.}
\end{figure}


\subsubsection{Computational Overhead} 
According to the results shown in Fig. \ref{fig:7-CapAc_exec_time}, the average total delay time caused by the BlendCAC operation of retrieving data from the client to server is 250 ms on satellites. Since the ground sites have much more computation capacity than the satellites, the execution time of the whole data querying process on the ground communication is about 60 ms. The total delay includes the round trip time (RTT), time for querying capability data from the smart contract, time for parsing JSON data from the request, and time for identity authentication and the access right validation. The token processing task is mainly responsible for fetching token data from the smart contract and introduces the highest workload among the AC operation stages. As the most computing intensive stage, the execution time of token processing is about 60 ms on the satellite, and the same operation on the ground communication only needs 10 ms.

The entire AC process is divided into two steps, identity authentication and capability token verification. Since the identity authentication process needs to interact with smart contract twice for querying VZone and Vnode data separately, identity authentication processing time is 152 ms, which is almost twice as much as that of execution on capability-based access control stage: 63 ms. The execution time of the AC process is about 214.5 ms (152 + 62.5) on satellites, which accounts for almost 86\% of the entire data service processing time.

\subsubsection{Communication Overhead}
Due to the high overhead introduced by querying token data from the smart contract in token processing stage, a token data caching solution is introduced in the BlendCAC system to reduce the network latency. When the client sends a service request to the server, the service provider extracts cached token data from the local storage to valid authorization. The service providers regularly update cached token data by checking smart contract status. The token synchronization time is in consistence with the block generation time, which is about 15 seconds in the Ethereum blockchain network. Simulating a regular service request allows us to measure how long it takes for the client to send a request and retrieve the data from the server. 

\begin{figure} [t]
\begin{center}
\begin{tabular}{c} 
\includegraphics[height=8.5cm]{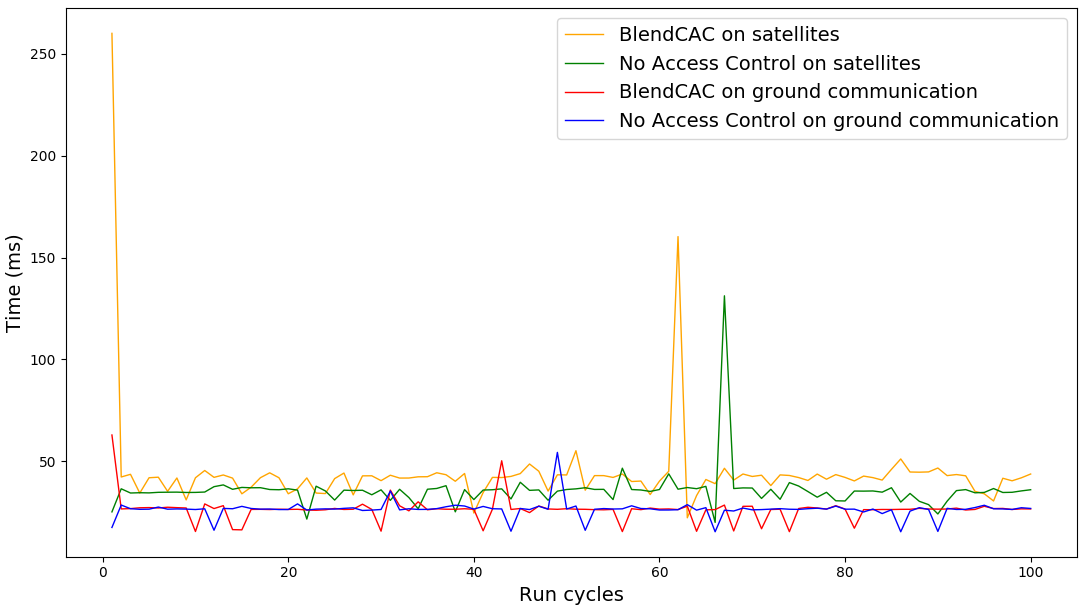}
\end{tabular}
\end{center}
\caption[example] { \label{fig:8-CapACVsNoCapAC} SATCOM Latency of BlendCAC.}
\vspace{-10pt}
\end{figure}

Figure \ref{fig:8-CapACVsNoCapAC} shows the overall SATCOM communication latency incurred and compares the execution time of the BlendCAC and a benchmark without any AC enforcement (BwoAC). At the beginning, a long SATCOM delay is observed in the first service request scenario, in which the service provider communicated with the smart contract and cached the token data. However, by processing the local cached token data for authorization validation, the SATCOM latency decreases quickly and becomes stable during the subsequent service requests. At the satellite device, the benchmark without AC enforcement (BwoAC) takes an average of 35 ms for fetching requested data versus the BlendCAC that has an average of 44 ms. Te results demonstrate that the proposed BlendCAC scheme only introduces about 9 ms extra latency. The overhead in terms of delay by the AC enforcement is even more trivial on ground communication nodes. The average time for querying data with AC is about 26 ms, which is almost the same as the average time of querying data without AC. However, the benefit of the secure BlendCAC outweighs the small latency cost. 

\subsubsection{Processing Overhead}
The delegation certificate and capability token could be validated only if the related transactions to the smart contract have been approved by miners and recorded to new blocks. The transaction rate is proportional to the block generation time, which refers to the time consumed by the miners to verify new blocks. Table \ref{tab:bc_cost} demonstrates the impacts of the number of miners on the blockchain network as well as estimated financial cost for transactions. In each scenario, sixty blocks are appended to the blockchain and the average block generation time is calculated. In initial, only two miners run consensus algorithm. As more miners perform the proof of work, the block generation time drops down and finally becomes stable. As shown by Table \ref{tab:bc_cost}, the optimal number of miners to get minimum block generation time is 6 in our private blockchain network. As a central part of the Ethereum network, gas is used to pay for the computing resources consumed by miners. To evaluate process overhead resulting from gas cost, 100 transactions that assign delegate certificate and capability are created on the blockchain network, and the average gas cost for each transaction is 169576.15 Wei (in Ethereum, 1 ether=1.0 ${\times}$ 10$^{18}$ Wei), which is around \$0.22 considered ETC value during the writing of this paper (September 20, 2018) is 1 ETC = 212.77\$.

\begin{table}[ht]
\caption{Impact of block generation and financial cost.} 
\label{tab:bc_cost}
\begin{center}       
\begin{tabular}{|p{4.1cm}|p{1.4cm}|p{1.4cm}|p{1.5cm}|p{1.5cm}|p{1.4cm}|p{1.4cm}|} 
\hline\hline\hline
\multicolumn{7}{|c|}{\textbf{Time consumption of block  generation}} \\
\hline
\textbf{Number of miners} & 2 & 3 & 4 & 5 & 6 & 7 \\
\hline
\textbf{Time (ms)} & 16.07 & 15.65 & 13.58 & 9.37 & 7.73 & 7.95 \\
\hline\hline
\multicolumn{7}{|c|}{\textbf{Estimated financial cost of transaction}} \\
\hline
\textbf{Gas (Wei) } & \multicolumn{2}{|l|}{159,544.25} & 
\multicolumn{2}{|l|}{\textbf{ETC Price (USD)}} & \multicolumn{2}{|l|}{212.77}\\
\hline
\textbf{Transaction fee (ETC) } & \multicolumn{2}{|l|}{0.0010853} & 
\multicolumn{2}{|l|}{\textbf{Transaction fee (USD)}} & \multicolumn{2}{|l|}{0.22}\\
\hline\hline\hline
\end{tabular}
\end{center}
\end{table}

\subsection{Discussion}
The experimental results demonstrate that the proposed BlendCAC strategy is effective and efficient to protecting the space devices and services from an unauthorized access request. Compared to centralized AC solutions, the BlendCAC scheme has the following advantages:

\begin{itemize}

\item \emph{Decentralized Architecture}: Due to the decentralization provided by the blockchain technique, the proposed BlendCAC scheme allows masters to control their devices and resources instead of depending on a centralized third authority to establish the trust relationship with unknown nodes; thus, the bottleneck effect and the risk of malfunction resulting from centralized architecture are removed. Even in the worst case that a master is out of service, it has limited impact on the authentication (apart from adding new nodes to a virtual trust zone). 

\item \emph{Edge Computing Driven Intelligence}: Thanks to federated delegation mechanism and blockchain technology, the BlendCAC framework provides a device-driven AC strategy that is suitable for the distributed nature of a space communications network. Through transferring power and intelligence from the centralized cloud server to the space network edge, smart objects are capable of protecting their own resources, enforcing privacy, and securing user-defined data content, which is meaningful to distributive, scalable, heterogeneous, and dynamic space scenarios;

\item \emph{Fine Granularity}: In the BlendCAC system, each entity uses its unique block-chain address for identity authentication and joins the virtual trust zone, and a capability token is only assigned to the authenticated entity. It is difficult for attackers to access services by using fake identities. Enforcing access right validation on local service providers empowers those smart devices to decide whether or not to grant access to certain services according to the local environmental conditions. Fine-grained AC with lease privilege access principle prevents privilege escalation, even if an attacker steals capability token; and

\item \emph{Lightweight}: Compared to XML-based language for AC, such as XACML, JSON is a lightweight technology that is suitable for resource constrained platforms. Given the experimental results, our JSON based capability token structure introduces small overhead on the general performance.
\end{itemize}

Although the proposed BlendCAC mechanism has demonstrated these attractive features, using blockchain to enforce AC policy in space systems, it also incurs new challenges in performance and security. The transaction rate is associated with confirmation time of the blockchain data, which depends on the block size and the time interval between the generations of new blocks. Thus, the latency for transaction validation may not be able to meet the requirement in real-time SSA scenarios. In addition, as the amount of transactions increases, the blockchain becomes large. The continuously growing data introduces more overhead on storage and computing resources of each client, especially for resource constrained devices. Furthermore, the blockchain is susceptible to majority attack (also known as 51\% attacks), in which once an attacker takes over 51\% computing power of network by colluding selfish miners, they are able to control the blockchain and reverse the transactions. Finally, since the blockchain data is open to all nodes joined the blockchain network, such a property of transparency inevitably brings privacy leakage concerns. More research efforts are necessary to improve the trade-off when applying the BlendCAC in practical scenarios.

\section{Conclusions}
\label{sec:conclusion}
In this paper, we proposed a partially decentralized capability-based AC framework leveraging the smart contract and blockchain technology, called BlendCAC, to handle the challenges in AC strategies for SSA applications. A concept-proof prototype has been built in a SSA emulated physical network environment to verify the feasibility of the proposed BlendCAC scheme. The identity authentication scheme and Capability-based access model are transcoded to smart contracts and work on the private Ethereum blockchain network. The desktops and laptops serve as miners to maintain the sanctity of transactions recorded on the blockchain, while Raspberry PI devices act as edge computing nodes to access and to provide services. Extensive experimental studies have been conducted using a space network emulator and the results are encouraging. It validated that the BlendCAC scheme was able to efficiently and effectively enforce AC authorization and validation in a distributed and trustless network. This work has demonstrated that our proposed BlendCAC framework is a promising approach to provide a scalable, fine-grained and lightweight access control for space network applications.

While the reported work has shown significant potential, there is still a long way towards a complete decentralized security solution for real-world space scenarios. Deeper insights are expected. Part of our on-going effort is focused on further exploration of the blockchain based AC scheme for real-world space imagery access scenarios. Furthermore, designing a efficient consensus mechanism to address current issues in blockchain, such as lower transaction rate and 51\% attack, is another effort to enhance the blockchain network towards reliable SSA applications of RSO tracking and space weather monitoring.


\bibliography{report}   
\bibliographystyle{spiejour}   





\end{spacing}
\end{document}